\documentclass[twoside]{articlek}

\renewcommand{\refname}{REFERENCES}
\usepackage{nccmath}
\usepackage{amsmath}
\usepackage{booktabs}
\usepackage{url}
\def\be{\begin{equation}}
\def\ee{\end{equation}}

 \renewcommand\d{\mathrm{d}}			
\newcommand\der[2]{\frac{\d #1}{\d #2}}	
\newcommand\dder[2]{\frac{\d^2 #1}{\d #2^2}}	

\renewcommand\theta{\vartheta}
\newcommand\at[2]{\left.#1\right|_{#2}}
\newcommand\eu{\mathrm{e}}
\newcommand\C{^\circ\mskip-2mu\mathrm{C}}

\def\BibTeX{{\rm B\kern-.05em{\sc i\kern-.025em b}\kern-.08em
            T\kern-.1667em\lower.7ex\hbox{E}\kern-.125emX}}

\usepackage{graphicx}
\begin{document}
\sloppy
\twocolumn[{
{\large\bf MEASUREMENT AND SIMULATION OF A TEMPERATURE FIELD}\\

{\small S. Jankov\'{y}ch, samueljankovych@gmail.com, Gymn\'{a}zium Tilgnerova, Bratislava and P. Bokes, peter.bokes@stuba.sk, 
\'{U}JFI FEI Slovak University of Technology, Bratislava}\\


{\bf ABSTRACT.} 
Measurements of surface temperature fields are used to determine the heat transfer by conduction and convection from an inhomogeneously 
heated metallic tube into environment. For most of the here reported measurements
we use a low-cost infra-red camera Seek Thermal, but the results are also compared to measurements done with 
a professional Fluke Ti20 Thermal imager. The results are interpreted using two simple theoretical models which give estimates of
the thermal conductivity of the tube and the heat transfer coefficient between the tube and the environment. 
\\
}]
%

\section{MOTIVATION AND THE SETUP}

The main motivation for our work is to test capabilities and possible usage of a low-cost compact infra-red camera 
Seek Thermal Compact~\cite{Seek}. This camera can be easily attached to and used by any Android or iOS mobile device. 
Unfortunately, the supplied software exports only graphic images (JPG) so an additional 
post-processing had to be done using the software library matplotlib~\cite{matplotlib} and our own python program. 
Using a linear color-to-temperature transformation the black-and-with thermal image was processed into a temperature array and 
the program's output was either average temperature over a selected rectangle or individual line-scans of the temperature along the imaged surface, 
averaged over the perpendicular direction.
\begin{figure} [ht]  
                  
\begin{center}                        
\includegraphics[width=80mm]{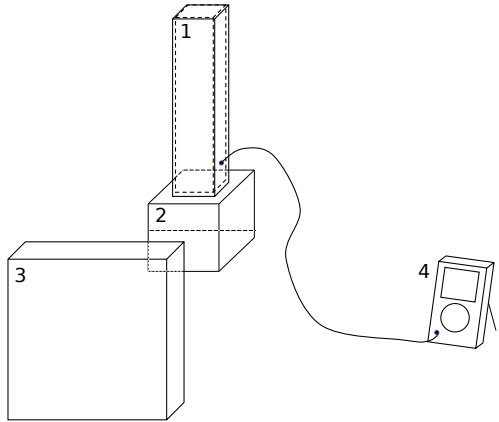}
    
\end{center}                         
\vspace{-2mm} 
\caption{The experimental setup.}
\label{fig:model}
\end{figure}  

In the Fig. \ref{fig:model} is the actual setup of the experiment. The hollow tube (1) is made of steel, sealed from the bottom 
and placed on a plastic pad (2). The tube had length $L=22.75$~cm and a square cross-section  with its edge length $a=3.6$~cm and its wall 
thickness $h=2.5$~mm. Its mass is $m=510$~g and its specific heat capacity is taken as $c_p = 502$ J/(kg$\cdot$K) (construction steel). 
White paint was sprayed on the tube's surface to achieve high enough emissivity. 
Between the pad and the tube we placed a thin resistive Ni spiral wire with resistance $R=1.3 \Omega$. 
During the heating, the power source was set to  constant voltage $U=4.5$~V and current $I=2.25$~A, leading to the total incoming heat 
flux $P =10.125$~W. We neglected the losses of the heat into the pad, which had much lower thermal conductivity than the metallic tube.
The ambient temperature was $\theta_\infty = 25 \C$.


\section{THEORETICAL MODELS}

We use two regimes to characterize the heat transfer processes. The first one is a stationary 
heat transfer where the surface temperature depends only on the vertical spatial variable - $x$ due to the symmetry of the setup. 
The differential equation describing this temperature profile is~\cite{Carslaw}
\be
 \lambda \dder{\theta}{x} - \frac{\alpha_1}{h} (\theta - \theta_{\infty})=0
\ee
with the boundary conditions (\ref{fig:model}),
\be
\theta(0)=\theta_0 , \quad
-\lambda \at{\der{\theta}{x}}{x=L}=\alpha_2 (\theta(L) - \theta_{\infty}).
\ee 
The amount of heat transferred depends on the temperature difference between tube's surface and ambient air ($\theta_{\infty}$). 
The heat transfer coefficient is different for the wall of the tube ($\alpha_1$) and at its top edge ($\alpha_2$). 
$h$ is the thickness of the wall and $L$ is the overall height of the tube. 
While the constant flux boundary condition at $x=0$ would be more appropriate, we have chosen the constant temperature instead, since
the temperature at this point ($\theta_0$) was also measured with a thermocouple.

The solution of the above equation is:
\be
\theta(x)=C \sinh{\left(\sqrt{\frac{\alpha_1}{h\lambda}}x\right)}
+(\theta_0-\theta_{\infty})\eu^{-\sqrt{\frac{\alpha_1}{h\lambda}}x} + \theta_{\infty}
\label{eqn1}
\ee
\be
C=\frac{(\sqrt{\frac{\alpha_1}{h\lambda}}-\frac{\alpha_2}{\lambda})(\theta_0-\theta_{\infty})\eu^{-\sqrt{\frac{\alpha_1}{h\lambda}}L}}{(\sqrt{\frac{\alpha_1}{h\lambda}}+\frac{\alpha_2}{\lambda})\sinh{(\sqrt{\frac{\alpha_1}{h\lambda}}L)}}
\ee
\\

The second regime is cooling of the tube once the heat source is switched off. A simple model 
that can be used to describe the relaxation of the average temperature is given by the energy balance,
\be
m c_p \der{\theta_{\textrm{av}}(t)}{t}= -S\alpha_1 (\theta_{\textrm{av}}(t)-\theta_{\infty})
\ee
with the initial condition
$\theta_{\textrm{av}}(0)=\theta_{\textrm{av},0}$.
It's solution is 
\be
\theta_{\textrm{av}}(t) = (\theta_{\textrm{av},0}-\theta_{\infty}) e^{-\frac{S\alpha_1}{m c_p} t} +\theta_{\infty}
\label{eqn2}
\ee
where $\theta_{\textrm{av}}$ is the average temperature of the system. $\theta_{\textrm{av},0}$ represents 
the average temperature at which we turn of heating of the system. The cooling depends on the surface of the tube $S$ 
through which the heat escapes into environment and $m$ and $c_p$ are the mass and the specific heat capacity 
of the tube. 

\section{RESULTS AND DISCUSSION}

The tube is first heated with a constant power flux from the heat source. The steady state is obtained after about 30 minutes, when the temperature 
of the thermocouple started to fluctuate around a stable value. The magnitude of these fluctuations was $\pm 0.8 \C$.
The resulting vertical temperature profile of the tube has been obtained from the thermal picture and saved for processing. 

\begin{figure} [ht]                     
\begin{center}                  
\includegraphics[width=80mm]{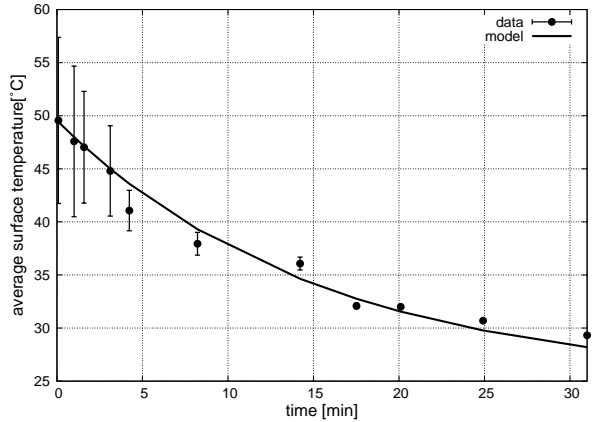}   
\end{center}                         
\vspace{-2mm} \caption{Comparison of fitted data versus measured data from infra-red camera with average temperature as a function of time. Error bars represent interval in which local surface temperature changes within that picture}
\label{fig:tdep} 
\end{figure}

Next, the power supply was switched off and we took 12 thermal pictures with variable time step within a total time interval of 30 minutes. 
For each picture we calculated the overall average surface temperature which resulted in the data points shown in Fig.\ref{fig:tdep}. 
However, since the temperature was at the initial times very inhomogeneous, we show the extremal values of the local surface temperature as errorbars.
Fitting the experimental data to the model in the Eqn.~\ref{eqn2} we obtain the estimate of the heat transfer coefficient for the tube's wall
$\alpha_1= 9.14 \pm 0.49$ W/(m$^2 \cdot$K). The regression coefficient of this fit is $R^2 = 0.9761$.

\begin{figure} [ht]                     
\begin{center}                       
\includegraphics[width=80mm]{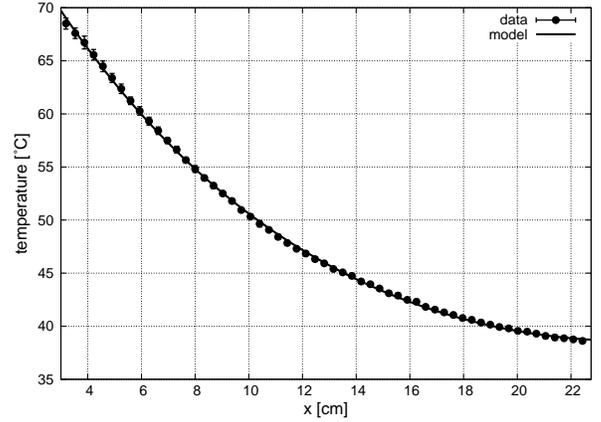}   
\end{center}                         
\vspace{-2mm} \caption{Comparison of fitted data versus measured data from infra-red camera with temperature as a function of position on the cylinder ($x$)}
\label{fig:xdep}
\end{figure}

Having determined the coefficient $\alpha_1$, we can return to the analysis of the steady-state profile (\ref{fig:xdep}). The measured data points 
are fitted to the model Eq.~\ref{eqn1}. The result of the fitting procedure are estimates of the heat conductivity 
$\lambda = 45.0 \pm 2.4$ W/(m$\cdot$K), and the heat transfer coefficient of the tube's edge $\alpha_2 = 65.4 \pm 5.3$ W/(m$^2 \cdot$K).
We have found that also $\theta_0$ had to be also fitted, instead of using the measured value of the thermocouple positioned approximately 
at $x=0$. The values of the latter were only $\theta_\textrm{TC} = 64 \C$. 

For comparison we made analogous measurements with a professional camera Fluke Ti20 Thermal imager. The resulting values for the parameters
are in satisfactory agreement with the former ones: 
$\alpha_1= 9.2 \pm 0.5$ W/(m$^2 \cdot$K) with much better fit $R^2=0.9817$ and $\lambda = 49.5 \pm 3.5$ W/(m$\cdot$K) and $\alpha_2 = 73.1 \pm 20.4$ W/(m$^2 \cdot$K).

\section{CONCLUSIONS}
The Seek Thermal Compact camera is a good low-cost thermal imaging camera. 
We have shown that with this camera can be used for physics measurements and engineering applications. 
The drawback is that unlike professional cameras the supplied software does not support more detailed analysis of the temperature field.

\section{ACKNOWLEDGMENTS}
We thank BEZ Transformatory, a.s. for lending the Fluke Ti20 thermal camera. This work was supported by the grant KEGA 002STU-4/2019 ``Heat transfer 2021''.

\end{document}